\documentclass[fp]{jpsj2}
\usepackage{txfonts}
\usepackage{epsf}
\usepackage{graphicx}
\usepackage{epstopdf}
\usepackage{multirow}
\usepackage{booktabs}
\usepackage{amsfonts}
\usepackage{dcolumn}%
\usepackage{color,ulem}
\usepackage{bm}
\usepackage{calligra}
\usepackage{calrsfs}
\usepackage{mathrsfs}
\usepackage{color}
\usepackage[maxfloats=256]{morefloats}
\usepackage{latexrelease}

\begin{document}

\title{Perturbative Analysis of the Field-Free Josephson Diode Effect in a Multilayered Josephson Junction
}
\author{Shin-ichi Hikino$^{*}$}
\inst{National Institute of Technology (KOSEN), Fukui College, Sabae, Fukui 916-8507, Japan}
\vspace{10pt}
\date{\today}

\abst{
The Josephson diode effect (JDE) is a novel phenomenon in which a superconducting junction exhibits asymmetric Josephson currents with respect to the superconducting phase difference.
In this study, we theoretically investigate how the interplay between a static exchange field and Rashba spin-orbit interaction (RSOI) influences the JDE.
By employing the quasiclassical Green's function method and perturbative calculations, we derive analytical expressions for the Josephson current in a junction composed of a ferromagnetic layer and a normal metal with RSOI.
Remarkably, the JDE is found to emerge even in the absence of any external magnetic field.
In this regime, the Josephson current is exclusively carried by spin-triplet Cooper pairs, as spin-singlet  components are strongly suppressed by the ferromagnet.
Furthermore, our results show that the efficiency of the JDE can be enhanced by tuning the thickness of the normal metal and the strength of the RSOI.
These findings offer valuable theoretical guidelines for the design of superconducting devices exhibiting nonreciprocal transport effects.
}


\maketitle 


\section{Introduction}
The remarkable advancements in modern science and technology owe much to the development of semiconductors. 
Semiconductor devices such as diodes, which exhibit non-reciprocal transport properties, are essential components of integrated circuits~\cite{sze}. 
These circuits are widely used in smartphones, computers, automobiles, medical equipment, and various other advanced technologies. 
Research and development of these semiconductor devices continue to this day, driving further progress in modern society.

Recently, non-reciprocal transport has been experimentally observed in superconductors (Ss)\cite{wakatsuki-sa3, ando-nature584, narita-nt17, lyu-nc12, shin-prr5, lin-np18, hou-prl131, sundaresh-nc14, satchell-jap133} 
and theoretically analyzed using both microscopic and phenomenological approaches\cite{tokura-nc9, daido-prl128, he-njp24, ilic-prl128, yuan-nas119}.
This non-reciprocal phenomenon arises from an asymmetry between the forward supercurrent ($I_{\rm c+}$) 
and the backward supercurrent ($I_{\rm c-}$), leading to $|I_{\rm c+}| \neq |I_{\rm c-}|$.
This dissipationless non-reciprocal transport is referred to as the superconducting diode effect (SDE).
The SDE holds promise for the development of low-power electronic devices, as it operates without dissipation, 
in contrast to conventional diode effects based on semiconductors. 

In addition to the superconducting diode effect (SDE) observed in Ss, 
Josephson junctions---composed of two Ss separated by non-superconducting materials~\cite{josephson, degennes, likhalev, golubov-rmp, buzdin-rmp, bergeret-rmp}---also exhibit a diode effect, known as the Josephson diode effect (JDE), 
under certain conditions~\cite{chen-prb98, misaki-prb103, kopasov-prb103, pal-nt17, wu-nature604, davydova-sa8, souto-prl129, kokkeler-prb106, zhang-prx12, ciaccia-prr5, tian-apl123, gupta-nc14, greco-apl123, chiles-nlt23, legg-prb108, wei-prb108, hu-prl130, lu-prl131, steiner-prl130, cheng-prb107, sun-prb108, kim-nc15, li-acs18, zhang-pra21, coraiola-acs18, huang-apl125, fracassi-apl124, roig-prb109, debnath-prb109, zhang-prb109, mori-arxiv}. 
The key requirements for realizing the JDE are inversion symmetry breaking (ISB) and time-reversal symmetry breaking (TRSB). 
ISB and TRSB can be readily achieved in materials that exhibit, for instance, Rashba spin-orbit interaction (RSOI) and are subjected to an external magnetic field, respectively. 
Supporting the theoretical prediction~\cite{souto-prl129}, the JDE has been experimentally demonstrated using a SQUID setup~\cite{ciaccia-prr5, greco-apl123}.
In these experiments, the ISB is introduced via an artificially asymmetric geometry of the Josephson junction, enabling an asymmetric Josephson current.
the TRSB is achieved by applying an external magnetic field to the SQUID loop. 
Moreover, the JDE without an external magnetic field, i.e., the field-free JDE, has also been theoretically predicted~\cite{kokkeler-prb106, hu-prl130, 
lu-prl131, wei-prb108} and experimentally observed~\cite{wu-nature604, chiles-nlt23, zhang-pra21, coraiola-acs18}. 
For instance, the ISB and the TRSB were introduced via asymmetric tunneling barriers and electrical magnetochiral anisotropy, respectively~\cite{li-acs18}. 
Other mechanisms to achieve ISB and TRSB include using spin-orbit coupled materials or multi-terminal junctions, among others~\cite{wu-nature604, chiles-nlt23, zhang-pra21, coraiola-acs18, kokkeler-prb106, hu-prl130, lu-prl131, wei-prb108}, but detailed discussions are beyond the scope of this introduction.
It is also important to note that, in addition to ISB and TRSB, the presence of higher-harmonic components in the Josephson current is required to realize the JDE. 

Josephson junctions under ISB and TRSB can exhibit a finite phase shift in their ground-state phase difference, 
characterizing so-called $\varphi_0$-junction behavior~\cite{buzdin-prl1, goldobin-prl107, bergeret-epl, szombati-natp, mayer-natc, strambini, hikino-ph}. 
This anomalous phase shift not only signals the symmetry breaking but also underpins the appearance of nonreciprocal Josephson current flow, 
known as the Josephson diode effect (JDE).
It is important to note that the $\varphi_0$ phase shift is distinct from the $\pi$ phase shift observed 
in superconductor/ferromagnet/superconductor (S/F/S) junctions~\cite{golubov-rmp, buzdin-rmp, ryazanov-prl, kontos-prl}.
While the $\pi$ shift originates from the oscillatory behavior of Cooper pairs in the ferromagnet, 
the $\varphi_0$ shift arises due to the simultaneous breaking of inversion and time-reversal symmetries, 
typically involving spin-orbit coupling and magnetic interactions. 
It should be noted, however, that the presence of a $\varphi_0$ phase shift alone is not sufficient to realize the JDE.  
In $\varphi_0$-junctions with a purely sinusoidal current-phase relation (CPR), represented as $I(\theta) = I_{\rm c} \sin(\theta + \varphi_0)$,  
the forward and backward Josephson currents remain symmetric in magnitude.  
Therefore, such junctions do not exhibit the nonreciprocal behavior characteristic of the JDE.  
Here, $I_{\rm c}$ is the Josephson critical current, and $\theta$ is the superconducting phase difference between the two Ss.
To induce the JDE, it is essential that the CPR contains asymmetric components,  
which can arise from higher-harmonic contributions and additional cosine terms.  
A phenomenological form of the CPR, expected under ISB and TRSB, can be expressed as
\begin{equation}
I(\theta) = \sum_{n=1}^{\infty} \left[ I_{\rm c}^{(n)} \sin(n \theta + \varphi_0^{(n)}) + \tilde{I}_{\rm c}^{(n)} \cos(n \theta + \tilde{\varphi}_0^{(n)}) \right],
\label{gene}
\end{equation}
where $I_{\rm c}^{(n)}$ and $\tilde{I}_{\rm c}^{(n)}$ are the amplitudes of the $n$th-order sine and cosine harmonics, respectively,  
and $\varphi_0^{(n)}$, $\tilde{\varphi}_0^{(n)}$ denote their respective phase shifts.
This generalized CPR no longer satisfies $I(\theta) = -I(-\theta)$ in general,  
and the critical currents in the forward and backward directions become unequal. 
Such asymmetry enables the emergence of the JDE even in $\varphi_0$-junctions.

Josephson junctions incorporating magnetic materials exhibit an intriguing phenomenon in addition to the $\varphi_0$ phase shift. 
Spin-triplet Cooper pairs (STCs) can be induced via the proximity effect, 
even when the superconductors are conventional $s$-wave superconductors~\cite{bergeret-rmp}. 
In particular, Josephson junctions consisting of two Ss separated by either a ferromagnet with non-uniform magnetization~\cite{bergeret-prl86, fominov-prb75, alidoust-prb81, robinson-science} 
or a multilayer ferromagnet~\cite{volkov-prb90, bergeret-prb68, houzet-prb76, volkov-prb81, khaire} 
can generate STCs composed of equal-spin electron pairs with $|S|=1$, which cannot be induced in uniformly magnetized ferromagnets. 
A key characteristic of these STCs is that they are immune to the exchange field in ferromagnets, 
in contrast to spin-singlet Cooper pairs (SSCs), both of which are suppressed by the exchange field. 
However, it has been shown that the $\varphi_0$ phase shift in Josephson junctions containing multilayer ferromagnets and normal metals with RSOI (RM) is typically very small in the diffusive transport regime~\cite{hikino-ph}. 
This suggests that the efficiency of the JDE in such systems is also low. The small JDE arises because the Josephson current is primarily carried by spin-singlet Cooper pairs (SSCs), whereas the $\varphi_0$ shift originates from spin-triplet components. 
Therefore, if the Josephson current carried by SSCs is suppressed and the contribution from STCs becomes dominant, 
the efficiency of the JDE is expected to be significantly enhanced.

In this paper, we theoretically study the Josephson diode effect (JDE) in a Josephson junction with a metallic multilayer.
Based on the quasiclassical Green's function theory and perturbative calculations, 
we formulate both the first-harmonic Josephson current (FHJC) and the second-harmonic Josephson current (SHJC).
The Josephson current carried by spin-singlet Cooper pairs (SSCs) is negligible in the present setup, 
as we employ a thick ferromagnetic metal in the multilayer to suppress the SSC contribution.
The FHJC and SHJC exhibit damped oscillatory behavior as a function of the length of the normal metal with the Rashba spin-orbit interaction (RSOI), 
leading to sign changes in the Josephson current.
Moreover, the direction of the FHJC and SHJC can be tuned by adjusting the RSOI strength.
Remarkably, a sizable $\varphi_0$ phase shift appears in both the FHJC and the SHJC even in the absence of any external magnetic field, owing to the suppression of the SSCs and the interplay with the RSOI.
As a result, the JDE can emerge without applying an external magnetic field.
We also evaluate the efficiency of the JDE within the present perturbative framework.
Finally, we discuss potential guidelines to further enhance the JDE efficiency.

\section{First-harmonic Josephson current}\label{sec:formulation1}
\subsection{Setup and equations for the $S_{L}/F_{L}/F/RM/F_{R}/S_{R}$ junction} 
\begin{figure}[t]
\begin{center}
\vspace{10 mm} 
\includegraphics[width=11cm]{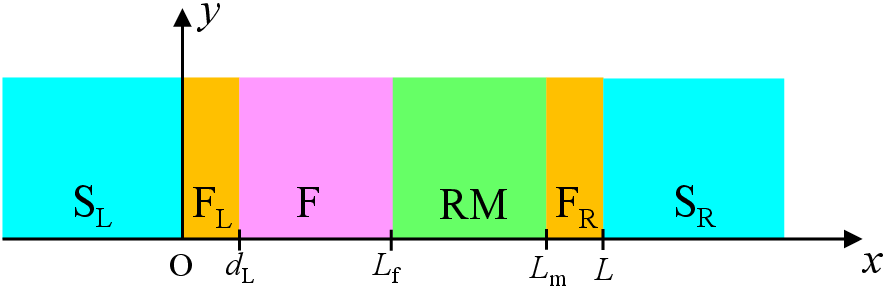}
\caption{ (Color online) 
The schematic diagram of the ${\rm S}_{\rm L}/{\rm F}_{\rm L}/{\rm F}/{\rm RM}/{\rm F}_{\rm R}/{\rm S}_{\rm R}$ junction is shown, 
where ${\rm S}_{\rm L}$ (${\rm S}_{\rm R}$) denotes the spin-singlet superconductor on the left (right), 
${\rm F}_{\rm L}$ (${\rm F}_{\rm R}$) the thin ferromagnetic metal on the left (right), ${\rm F}$ the thick ferromagnetic metal, 
and ${\rm RM}$ normal metal with the RSOI. 
The thick ferromagnet ${\rm F}$ serves as a barrier that suppresses the contribution of spin-singlet Cooper pairs. 
The total thicknesses of different regions are defined as $L_{\rm f} = d_{\rm L} + d_{\rm f}$, $L_{\rm m} = L_{\rm f} + d_{\rm m}$, and $L = L_{\rm m} + d_{\rm R}$, 
where $d_{\rm L(R)}$ are the thicknesses of ${\rm F}_{\rm L(R)}$, $d_{\rm f}$ that of ${\rm F}$, and $d_{\rm m}$ that of ${\rm RM}$.
}
\label{gm}
\end{center}
\end{figure}
As illustrated in Fig.~\ref{gm}, we consider a ${\rm S}_{\rm L}/{\rm F}_{\rm L}/{\rm F}/{\rm RM}/{\rm F}_{\rm R}/{\rm S}_{\rm R}$ junction,
where ${\rm S}_{\rm L}$ (${\rm S}_{\rm R}$) denotes the spin-singlet superconductor on the left (right),
${\rm F}_{\rm L}$ (${\rm F}_{\rm R}$) denotes the thin ferromagnetic metal on the left (right),
${\rm F}$ is a thick ferromagnetic metal, and ${\rm RM}$ is a normal metal with the RSOI.
The thick ferromagnet ${\rm F}$ acts as a barrier that suppresses the contribution of SSCs.
The total lengths of different regions are defined as $L_{\rm f} = d_{\rm L} + d_{\rm f}$, $L_{\rm m} = L_{\rm f} + d_{\rm m}$, and $L = L_{\rm m} + d_{\rm R}$,
where $d_{\rm L (R)}$ is the thickness of ${\rm F}_{\rm L}$ (${\rm F}_{\rm R}$), $d_{\rm f}$ is the thickness of ${\rm F}$, and $d_{\rm m}$ is the thickness of ${\rm RM}$.
In the following, we assume that the system's temperature is close to the superconducting transition temperature
and focus on the qualitative behavior of the Josephson current.

Assuming the system is in the diffusive transport region and near the superconducting transition temperature, 
we apply the linearized Usadel equation to the ${\rm S}_{\rm L}/{\rm F}_{\rm L}/{\rm F}/{\rm RM}/{\rm F}_{\rm R}/{\rm S}_{\rm R}$ junction. 
The linearized Usadel equation in the region $j$(=${\rm F}_{\rm L}$, F, RM, and ${\rm F}_{\rm R}$) is written as~\cite{bergeret-prl110} 
\begin{eqnarray}
i\hbar D\tilde \partial^{2} _x\hat f^{j}(x) - i2\hbar {{\omega}} \hat f ^{j}(x)
- h_{x}\left( x \right)\left[ {{{\hat \sigma }_x},\hat f^{j}(x)} \right] 
- h_{z}\left( x \right)\left[ {{{\hat \sigma }_z},\hat f^{j}(x)} \right] &=& \hat 0, 
\label{usadel-1} \nonumber \\
\end{eqnarray}
where
\begin{eqnarray}
{{\tilde \partial }_x} = \left\{ \begin{array}{l}
{\partial _x} \bullet  - i{\alpha _{\rm R}}\left[ {{{\hat \sigma }_y}, \bullet } \right],\,\,\,\,\,\,\,\,\,L_{\rm f} < x < L_{\rm m}\\
{\partial _x} \bullet \,\,\,\,\,\,\,\,\,\,\,\,\,\,\,\,\,\,\,\,\,\,\,\,\,\,\,\,\,\,\,\,,\,\,\,\,\,\,\,\,\,{\rm{other}}
\end{array} \right.. 
\end{eqnarray}
${\hat f}^{j}(x)$ is the anomalous Green's function, represented as a $2\times 2$ matrix.
$\alpha_{\rm R}$ is the RSOI constant in the RM. 
$D$ is the diffusion coefficient, which is assumed to have the same value in each region. 
$\omega = \pi k_{\rm B} T/\hbar$ is the fermion Matsubara frequency at $n = 0$. 
$\hat{\sigma}_{x (y,z)}$ is the $x (y,z)$ component of Pauli matrix. 
$[\hat A, \hat B]$ is the commutator. 

The exchange field ${\bm h}_{\rm ex} (x)$ in ferromagnetic metals is given by 
\begin{eqnarray}
{ {\bm h}_{{\rm{ex}}}}\left( x \right) &=& \left\{ \begin{array}{l}
h_{x}^{\rm L}\,{{\bm e}_x}+h_{z}^{\rm L}\,{{\bm e}_z},\,\,\,\,\,\,0 < x < {d_{\rm L}}\\
h\,{{\bm e}_z},\,\,\,\,\,\,\,\,\,\,\,\,\,\,\,\,\,\,\,\,\,\,\,\,\,\,{d_{\rm L}} < x < L_{\rm f}\\
0,\,\,\,\,\,\,\,\,\,\,\,\,\,\,\,\,\,\,\,\,\,\,\,\,\,\,\,\,\,\,\,\,{L_{\rm f}} < x < L_{\rm m}\\
h_{x}^{\rm R}\,{{\bm e}_x}+h_{z}^{\rm R}\,{{\bm e}_z},\,\,\,\,\,{L_{\rm m}} < x < L\\
\end{array} \right. ,
\end{eqnarray}
where $\bm{e}_{x(z)}$ is the unit vector of {\it x} ({\it z}) direction. 
$h_{x}^{\rm L (R)} = h \sin \theta_{\rm L (R)}$ and $h_{z}^{\rm L (R)} = h \cos \theta_{\rm L (R)}$. 
$\theta_{\rm L (R)}$ is the polar angle of the magnetization in the ${\rm F}_{\rm L (R)}$. 

To obtain special solutions of Eq.~(\ref{usadel-1}), we apply suitable boundary conditions to general solutions of Eq.~(\ref{usadel-1}) as follows~\cite{demler-prb}, 
\begin{eqnarray}
\hat f^{\rm S_{\rm L}}(x=0) &=& \hat f^{\rm F_{\rm L}}(x=0), 
\label{bc1}\\
\hat f^{\rm F_{\rm L}}(x=d_{\rm L}) &=& \hat f^{\rm F }(x=d_{\rm L}), 
\label{bc2}\\
{\left. \frac{d}{dx} {\hat f}^{\rm F_{\rm L}} (x) \right|_{x=d_{\rm L}}} &=& {\left. \frac{d}{dx} {\hat f}^{\rm F} (x) \right|_{x=d_{\rm L}}}, 
\label{bc3}\\
{\hat f}^{\rm F} \left(x=L_{\rm f} \right) &=& {\hat f}^{\rm RM} \left(x=L_{\rm f} \right), 
\label{bc4} \\
{\left. \frac{d}{dx} {\hat f}^{\rm F} (x) \right|_{x=L_{\rm f}}} &=& {\left. \frac{d}{dx} {\hat f}^{\rm RM} (x) \right|_{x=L_{\rm f}}},
\label{bc5} \\
{\hat f}^{\rm RM} \left( x=L_{\rm m} \right) &=& {\hat f}^{\rm F_{\rm R}} \left( x=L_{\rm m} \right), 
\label{bc6} \\
{\left. \frac{d}{dx} {\hat f}^{\rm RM} (x) \right|_{x=L_{\rm m}}} &=& {\left. \frac{d}{dx} {\hat f}^{\rm F_{\rm R}} (x) \right|_{x=L_{\rm m}}}, 
\label{bc7} \\
\hat f^{\rm F_{\rm R}}(x=L) &=& \hat f^{\rm S_{\rm R}}(x=L). 
\label{bc8}
\end{eqnarray}
For simplicity, we assume that all regions have the same conductivity.

\subsection{Anomalous Green's functions in the RM}
The Josephson current flowing through the junction can be calculated by using the anomalous Green's function, 
which is given by 
\begin{eqnarray}
\begin{array}{l}
{{\hat f}^j}(x) = \left( {\begin{array}{*{20}{c}}
{f_{ \uparrow  \uparrow}^j\left( x \right)}&{f_{ \uparrow  \downarrow }^j\left( x \right)}\\
{f_{ \downarrow  \uparrow}^j\left( x \right)}&{f_{ \downarrow  \downarrow }^j\left( x \right)}
\end{array}} \right)
 = \left( {\begin{array}{*{20}{c}}
{ - f_{tx}^j\left( x \right) + if_{ty}^j\left( x \right)}&{f_{s}^j\left( x \right) + f_{tz}^j\left( x \right)}\\
{ - f_{s}^j\left( x \right) + f_{tz}^j\left( x \right)}&{f_{tx}^j\left( x \right) + if_{ty}^j\left( x \right)}
\end{array}} \right).
\end{array}
\end{eqnarray}
$f^{m}_{s}(x)$ is the anomalous Green's function of the SSC and 
whereas $f^{m}_{tx(ty)}(x)$ and $f^{m}_{tz}(x)$ are the anomalous Green's functions of the STC with $|S_{z}|=1$ and $|S_{z}|$=0, respectively. 

Within the rigid boundary condition~\cite{buzdin-rmp}, 
the anomalous Green's function of the ${\rm S}_{\rm L (R)}$ near the superconducting transition temperature is given by
%
\begin{eqnarray}
\hat f^{\rm S_{\rm L ({\rm R})}}(x)|_{x=0(L)} 
= - \hat \sigma_{y}
\frac{ \Delta_{\rm L(R)} }{ \hbar \omega }
\label{fs}, 
\end{eqnarray}
where $\Delta_{\rm L(R)}$ is the superconducting gap of the ${\rm S}_{\rm L (R)}$. 

Solving Eq.~(\ref{usadel-1}), we obtain the general solutions in the F and RM regions as follows, 
%
\begin{eqnarray}
\left( \begin{array}{l}
f_{s}^{{\rm{F}}}\left( x \right)\\
f_{tx}^{{\rm{F}}}\left( x \right)\\
f_{tz}^{{\rm{F}}}\left( x \right)
\end{array} \right) &=& 
A \left( \begin{array}{l}
0\\
1\\
0
\end{array} \right){e^{ x/\xi} } + 
B \left( \begin{array}{l}
0\\
1\\
0
\end{array} \right){e^{ -x/\xi} } \nonumber \\
&+&
C \left( \begin{array}{l}
1\\
0\\
-1
\end{array} \right){e^{\kappa_{+} x} } + D \left( \begin{array}{l}
1\\
0\\
-1
\end{array} \right){e^{ -\kappa_{+} x}} 
 + E \left( \begin{array}{l}
1\\
0\\
1
\end{array} \right){e^{ \kappa_{-} x }} 
 + F\left( \begin{array}{l} 
1\\
0\\
1
\end{array} \right){e^{ -\kappa_{-} x }} 
\label{gs-f}, \nonumber \\
\end{eqnarray}
and 
%
\begin{eqnarray}
\left( \begin{array}{l}
f_s^{{\rm{RM}}}\left( x \right)\\
f_{tx}^{{\rm{RM}}}\left( x \right)\\
f_{tz}^{{\rm{RM}}}\left( x \right)
\end{array} \right) &=& 
M \left( \begin{array}{l}
1\\
0\\
0
\end{array} \right){e^{ x/\xi }} + N \left( \begin{array}{l}
1\\
0\\
0
\end{array} \right){e^{ -x/\xi }} \nonumber \\
 &+& G\left( \begin{array}{l}
0\\
-i\\
i
\end{array} \right){e^{i 2\alpha_{\rm R} x}}{e^{ x/\xi }} + H\left( \begin{array}{l} 
0\\
-i\\
i
\end{array} \right){e^{i2\alpha_{\rm R} x}}{e^{ -x/\xi }} \nonumber \\
&+& I\left( \begin{array}{l}
0\\
i\\
i
\end{array} \right){e^{ - i2\alpha_{\rm R} x}}{e^{ x/\xi }} 
+ J\left( \begin{array}{l}
0\\
i\\
i
\end{array} \right){e^{ - i2\alpha_{\rm R} x}}{e^{ -x/\xi }} 
\label{gs-frm}. 
\end{eqnarray}
Where $\xi = \sqrt{\hbar D /2\pi k_{\rm B} T}$ and $\kappa_{\pm} = \sqrt{(\hbar \omega \pm i h)/\hbar D}$. 
It should be noticed that $f_{ty}^{\rm F (RM)} (x)$ is exactly zero because the exchange field does not have the $y$-component~\cite{houzet-prb76,hikino-prb}. 

To obtain the anomalous Green's function of the ${\rm F}_{\rm L (R)}$, we carry out the Taylor expansion with $x$ for ${\hat f}^{\rm F_{\rm L (R)}}(x)$ 
assuming that $d_{\rm L (R)}$ is much smaller than $ \xi_{\rm f} = \sqrt{\hbar D/h} $ and 
apply Eq.~(\ref{bc1}) - (\ref{bc3}) (Eq.~(\ref{bc6}) - (\ref{bc8})) to ${\hat f}^{\rm F_{\rm L (R)}}(x)$. 
As a result, ${\hat f}^{\rm F_{\rm L}}(x)$ and ${\hat f}^{\rm F_{\rm R}}(x)$ are expressed as 
%
\begin{eqnarray}
\hat f ^{\rm F_{\rm L}} (d_{\rm L}) & \approx &
d_{\rm L} \partial _x { \hat f ^{\rm F}(d_{\rm L}) } + \hat f^{\rm S_{ \rm L } }(0)
+ i \frac{ d_{\rm L}^{2} h_{x}^{\rm L} }{ 2 \hbar D }  \left[ {{{\hat \sigma }_x},\hat f^{\rm S_{ \rm L } }(0) } \right] 
+ i \frac{ d_{\rm L}^{2} h_{z}^{\rm L} }{ 2 \hbar D }  \left[ {{{\hat \sigma }_z},\hat f^{\rm S_{ \rm L } }(0)} \right] 
\label{gf-fl}, 
\end{eqnarray}
and 
%
\begin{eqnarray}
\hat f ^{\rm F_{\rm R}} (L_{\rm m}) & \approx &
-d_{\rm R} \partial _x { \hat f ^{\rm RM}(L_{\rm m}) } + \hat f^{\rm S_{ \rm R } }(L)
+ i \frac{ d_{\rm R}^{2} h_{x}^{\rm R} }{ 2 \hbar D }  \left[ {{{\hat \sigma }_x},\hat f^{\rm S_{ \rm R } }(L) } \right] 
+ i \frac{ d_{\rm R}^{2} h_{z}^{\rm R} }{ 2 \hbar D }  \left[ {{{\hat \sigma }_z},\hat f^{\rm S_{ \rm R } }(L)} \right] 
\label{gf-fr}. 
\end{eqnarray}
For details of the calculation, refer  to Refs~[68] and [73]. 

For simplicity in the calculation of ${\hat f}^{\rm RM}(x)$, we assume that $d_{\rm f}$ and $d_{\rm m}$ are much larger than $\xi$. 
Applying Eqs.~(\ref{bc4}) and (\ref{bc5}) to (\ref{gs-f}) and (\ref{gs-frm}), and then 
substituting Eqs.~(\ref{fs}), (\ref{gs-f}), and (\ref{gs-frm}) into Eqs.~(\ref{gf-fl}) and (\ref{gf-fr}), 
we obtain a system of simultaneous equations for determining the coefficients that satisfy the boundary conditions(see APPENDIX~\ref{app:coef}). 
By solving this system of simultaneous equations, as shown in the APPENDIX~\ref{app:coef}, ${\hat f}^{\rm RM}(x)$ can be approximately expressed as follows,
%
%
\begin{equation}
f_{s}^{\rm RM}(x) \approx i \frac{\Delta_{\rm R}}{ \hbar \omega } e^{(x-L_{\rm m})/\xi}
\label{fs-rm},
\end{equation}
\begin{eqnarray}
f_{tx}^{\rm RM}(x) &\approx&
	\frac{2 D_{x}^{\rm L}}{ 1+\kappa_{\alpha}^{+} \xi }
	\left[
	e^{- 2 d_{\rm m}/\xi} e^{\kappa_{\alpha}^{-} \left(x-d_{\rm f} \right) }
	-e^{- \kappa_{\alpha}^{+} \left(x-d_{\rm f} \right)}
	\right]
	e^{-d_{\rm f}/\xi} \frac{\Delta_{\rm L}}{ \hbar \omega } \nonumber \\
	&+&
	\left[
	\frac{\alpha_{\rm R}^{-}}{ 2 } \left(D_{x}^{\rm R} - D_{z}^{\rm R} \right) e^{-\kappa_{\alpha}^{-} \left(x-L_{\rm m} \right)} 
	-
	\frac{\alpha_{\rm R}^{+}}{ 2 } \left(D_{x}^{\rm R} + D_{z}^{\rm R} \right) e^{-\kappa_{\alpha}^{+} \left(x-L_{\rm m} \right)} 
	\right]
	\frac{ \Delta_{\rm R} }{ \hbar \omega } \nonumber \\
	&+&
	\frac{ \mathscr{D}_{xz}^{\rm R} }{1+\kappa_{\alpha}^{+}\xi  }
	\left[
	e^{-\kappa_{\alpha}^{+} \left(x-d_{\rm f} \right)} - e^{\kappa_{\alpha}^{-} \left(x-d_{\rm f} \right)} e^{-2 d_{\rm m}/\xi} 
	\right]
	\frac{\Delta_{\rm R}}{\hbar \omega}
\label{fx-rm}, \\
 f_{tz}^{\rm RM}(x) &\approx&
	\frac{2 D_{x}^{\rm L}}{ 1+\kappa_{\alpha}^{+} \xi }
	\left[
	e^{- 2 d_{\rm m}/\xi} e^{\kappa_{\alpha}^{-} \left(x-d_{\rm f} \right) }
	-e^{- \kappa_{\alpha}^{+} \left(x-d_{\rm f} \right)}
	\right]
	e^{-d_{\rm f}/\xi} \frac{\Delta_{\rm L}}{ \hbar \omega } \nonumber \\
	&-&
	\left[
	\frac{\alpha_{\rm R}^{+}}{ 2 } \left(D_{x}^{\rm R} + D_{z}^{\rm R} \right) e^{-\kappa_{\alpha}^{+} \left(x-L_{\rm m} \right)} 
	+
	\frac{\alpha_{\rm R}^{-}}{ 2 } \left(D_{x}^{\rm R} - D_{z}^{\rm R} \right) e^{-\kappa_{\alpha}^{-} \left(x-L_{\rm m} \right)} 
	\right]
	\frac{ \Delta_{\rm R} }{ \hbar \omega } \nonumber \\
	&+&
	\frac{ \mathscr{D}_{xz}^{\rm R} }{1+\kappa_{\alpha}^{+}\xi  }
	\left[
	e^{-\kappa_{\alpha}^{+} \left(x-d_{\rm f} \right)} - e^{\kappa_{\alpha}^{-} \left(x-d_{\rm f} \right)} e^{-2 d_{\rm m}/\xi} 
	\right]
	\frac{\Delta_{\rm R}}{\hbar \omega}
\label{fz-rm},
\end{eqnarray}
and
\begin{eqnarray}
\mathscr{D}_{xz}^{\rm R} = 
i D_{x}^{\rm R}{\rm Im}
\left[
\alpha_{\rm R}^{+} \left(1 + \kappa_{\alpha}^{+} \xi \right) e^{\kappa_{\alpha}^{+}d_{\rm m}}
\right]
+
D_{z}^{\rm R}{\rm Re}
\left[
\alpha_{\rm R}^{+} \left(1 + \kappa_{\alpha}^{+} \xi \right) e^{\kappa_{\alpha}^{+}d_{\rm m}}
\right]. 
\end{eqnarray}
Where $\kappa_{\alpha}^{\pm} = 1/\xi \pm i 2 \alpha_{\rm R}$, 
$D_{x (z)}^{\rm L (R)} = d_{\rm L (R)}^{2} h_{x (z)}^{\rm L (R)}/\hbar D$, and $\alpha_{\rm R}^{\pm}=1 \pm i 2\alpha_{\rm R}d_{\rm R}$. 
In the next subsection, we calculate the first harmonic Josephson current (FHJC) using Eqs.~(\ref{fs-rm}) -- (\ref{fz-rm}).
\subsection{Formulation of the first harmonic Josephson current}
In the quasiclassical theory near the superconducting transition temperature, 
the FHJC in a system with the RSOI by considering $f_{ty}(x) =0$ can be obtained by 
\begin{eqnarray}
j_{\rm Q}(x) &=&
	i \frac{ e \pi D N_{\rm F} k_{\rm B} T }{ 2 }
	{\rm tr}
	\left[
	{\hat f}(x) \tilde \partial _x {\hat f}^{\dagger}(x) - {\hat f}^{\dagger}(x) \tilde \partial _x {\hat f}(x)
	\right]
	= j_{1}(x) + j_{2}(x)
\label{jf}, \\
j_{1}(x) &=&
	2e\pi D N_{\rm F} k_{\rm B} T
	{\rm Im}
	\left[
	f_{tx}(x) \partial_x {\hat f}_{tx}^{\dagger}(x)
	+f_{tz}(x) \partial_x {\hat f}_{tz}^{\dagger}(x)
	-f_{s}(x) \partial_x {\hat f}_{s}^{\dagger}(x)
	\right]
\label{jqf}, 
\end{eqnarray}
and 
\begin{eqnarray}
j_{2}(x) &=&
	-8\alpha_{\rm R}e\pi D N_{\rm F} k_{\rm B} T {\rm Im}
	\left[
	f_{tx}(x) f_{tz}^{\dagger}(x)
	\right]
\label{jaf}.
\end{eqnarray}
Here, $N_{\rm F}$ is the density of states per unit volume and per electron spin at the Fermi energy. 
The FHJC consists of two contributions in the presence of the RSOI. 
One is $j_{1}(x)$, which is non-zero in Josephson junctions without the RSOI. 
The other is $j_{2}(x)$, which is non-zero in Josephson junctions with the RSOI. 
It follows from Eqs.~(\ref{fs-rm}), (\ref{jqf}), and (\ref{jaf}) that the FHJC is solely carried by STCs. 
By substituting \( x = L_{\rm m} \) into Eqs.~(\ref{fx-rm}) and (\ref{fz-rm}), and then inserting the results into Eqs.~(\ref{jqf}) and (\ref{jaf}),  
the FHJC as a function of $d_{\rm f}$ and $d_{\rm m}$ is represented by 
%
\begin{eqnarray}
j_{1}(L_{\rm m}) &\approx& 
	-8\frac{ e \pi D N_{\rm F} k_{\rm B} T }{ \xi }
	\left(
	\frac{ \Delta }{\hbar \omega}
	\right)^{2}
	\left(
	a_{1} + b_{1}
	\right)
	e^{-L_{\rm m}/\xi} \sin \theta \nonumber \\
	&+&
	8\frac{ e \pi D N_{\rm F} k_{\rm B} T }{ \xi }
	\left(
	\frac{ \Delta }{\hbar \omega}
	\right)^{2}
	\left(
	a_{2} - b_{2}
	\right)
	e^{-L_{\rm m}/\xi} \cos \theta
\label{jq}, 
\end{eqnarray}
\begin{eqnarray}
a_{1} &=&
	\left\{
	2 D_{x}^{\rm L} D_{x}^{\rm R} 
	\left[ \xi \alpha_{\rm R} - \sin^{2} \left( 2\alpha_{\rm R} d_{\rm m} \right) \right]
	+
	D_{x}^{\rm L} D_{z}^{\rm R} 
	\left[ \xi \alpha_{\rm R} - \sin \left( 4\alpha_{\rm R} d_{\rm m} \right) \right]
	\right\}
	\sin \left(2 \alpha_{\rm R} d_{\rm m} \right), \nonumber \\
a_{2} &=&
	\left\{
	D_{x}^{\rm L} D_{z}^{\rm R} 
	\left[ 1 + 2 \cos \left( 2\alpha_{\rm R} d_{\rm m} \right) \right]
	+
	D_{x}^{\rm L} D_{x}^{\rm R} 
	\sin \left( 4 \alpha_{\rm R} d_{\rm m} \right)
	\right\}
	\sin \left(2 \alpha_{\rm R} d_{\rm m} \right), \nonumber \\
b_{1} &=&
	D_{x}^{\rm L} D_{z}^{\rm R} 
	\left[
	\cos \left(2 \alpha_{\rm R} d_{\rm m} \right) + 2\xi \alpha_{\rm R} \sin \left(2\alpha_{\rm R} d_{\rm m} \right)
	\right], \nonumber \\
b_{2} &=&
	\frac{1}{2} \xi \alpha_{\rm R} D_{x}^{\rm L} \left( D_{x}^{\rm R} + D_{z}^{\rm R} \right)
	\left[
	\cos \left( 2\alpha_{\rm R} d_{\rm m} \right) +2\xi \alpha_{\rm R} \sin \left(2\alpha_{\rm R} d_{\rm m} \right)
	\right], \nonumber
\end{eqnarray}
\begin{eqnarray}
j_{2} (L_{\rm m}) &\approx&
	16\alpha_{\rm R} e\pi D N_{\rm F} k_{\rm B} T 
	\left(
	\frac{ \Delta }{\hbar \omega}
	\right)^{2}
	c_{1} \sin \left(2\alpha_{\rm R} d_{\rm m} \right) e^{-L_{\rm m}/\xi} \sin \theta \nonumber \\
	&+&
	16\alpha_{\rm R} e\pi D N_{\rm F} k_{\rm B} T 
	\left(
	\frac{ \Delta }{\hbar \omega}
	\right)^{2}
	c_{2} \sin \left(2\alpha_{\rm R} d_{\rm m} \right) e^{-L_{\rm m}/\xi} \cos \theta, 
\label{ja}
\end{eqnarray}
\begin{eqnarray}
c_{1} &=&
	\xi \alpha_{\rm R} D_{x}^{\rm L} \left(D_{x}^{\rm R} + D_{z}^{\rm R} \right), \nonumber 
\end{eqnarray}
and
\begin{eqnarray}
c_{2} &=&
	D_{x}^{\rm L} \left(D_{z}^{\rm R} - D_{x}^{\rm R} \right). \nonumber
\end{eqnarray}
From Eqs.~(\ref{jq}) and (\ref{ja}), it can be clearly seen that a $\varphi_0$ phase shift emerges
due to the additional cosine term in the Josephson current, even in the absence of an external magnetic field.
It should be noted that the coefficients in these equations vanish 
when either the RSOI or the thin ferromagnetic layers are absent, and thus the FHJC cannot flow.
Consequently, the JDE does not occur under these conditions.

\section{Second harmonic Josephson current}
\subsection{Equations for the $S_{L}/F_{L}/F/RM/F_{R}/S_{R}$ junction}
In this subsection, we derive the second harmonic Josephson current (SHJC) by using a perturbative calculation~\cite{richard-prl110}. 
To obtain the SHJC, we calculate the anomalous Green's function in the RM by solving the nonlinear Usadel equation as follows, 
\begin{eqnarray}
i \hbar D \tilde \partial _x 
\left[
\check {g}^{\rm RM}(x) \tilde \partial _x \check {g}^{\rm RM}(x)
\right]
-i \hbar \omega \check{\rho}_{z} \check {g}^{\rm RM}(x)
= \check{0}
\label{n-usadel1}, 
\end{eqnarray}
where $\check{g}^{\rm RM}(x)$ is a $4 \times 4$ matrix, and $\check{\rho}_{z}$ is the $z$-component of the $4 \times 4$ Pauli matrix.
Here, we apply the normalization condition $\check{g}(x)\check{g}(x) = \check{1}$ and introduce a suitable parameterization to solve Eq.~(\ref{n-usadel1}). 
As a result, the nonlinear Usadel equation, corresponding to the (1,1) component in the particle-hole space, is represented by 
\begin{eqnarray}
\tilde \partial _x 
	\left(
	\sqrt{1 - {\hat F}^{\rm RM}(x) {\hat F}^{\rm RM, \dagger}(x) } \tilde \partial _x {\hat F}^{\rm RM}(x) 
	-
	{\hat F}^{\rm RM}(x)\tilde \partial _x \sqrt{1 - {\hat F}^{\rm RM, \dagger}(x) {\hat F}^{\rm RM}(x)}
	\right) 
	-
	\frac{2D}{\omega}{\hat F}^{\rm RM}(x) 
	={\hat 0} 
\label{n-usadel2}.
\end{eqnarray}
To solve Eq.~(\ref{n-usadel2}) perturbatively, we expand $\hat{F}^{\rm RM}(x)$ around its normal-state solution.
As a result, $\hat{F}^{\rm RM}(x)$ can be expressed as 
\begin{equation}
{\hat F}^{\rm RM}(x) \approx {\hat f}^{\rm RM}(x) + \delta {\hat f}^{\rm RM}(x)
\label{f-apx}.
\end{equation}
Where ${\hat f}^{\rm RM}(x)$ is the solution of Eq.~({\ref{usadel-1}}). 
$\delta {\hat f}^{\rm RM}(x)$ is the perturbative term, from which we can obtain the SHJC. 
Substituting Eq.~(\ref{f-apx}) into Eq.~(\ref{n-usadel2}) and expanding square root parts in Eq.~(\ref{n-usadel2}), i.e., $\sqrt{1-x}$ with the small $x$, 
the linear Usadel equation to calculate the SHJC is expressed as 
%
\begin{eqnarray}
\tilde \partial^{2} _x \delta {\hat f}^{\rm RM}(x) - \frac{2 \omega}{D} \delta {\hat f}^{\rm RM}(x) &=&
	\frac{1}{2} 
	\tilde \partial _x 
	\left[
	{\hat f}^{\rm RM}(x) {\hat f}^{\rm RM, \dagger}(x) \tilde \partial _x {\hat f}^{\rm RM}(x)
	-
	{\hat f}^{\rm RM}(x) \tilde \partial _x 
	\left(
	{\hat f}^{\rm RM, \dagger}(x) {\hat f}^{\rm RM}(x)
	\right)
	\right]
\label{usadel3}. \nonumber \\
\end{eqnarray}
Assuming $\xi \alpha_{\rm R} \ll 1$ but $\alpha_{\rm R} \neq 0$ and $d_{\rm f (m)}/\xi \gg 1$, 
we substitute Eqs.~(\ref{fs-rm}) -– (\ref{fz-rm}) into Eq.(\ref{usadel3}),
and subsequently integrate Eq.~(\ref{usadel3}) with respect to $x$. 
As a result, the anomalous Green's function in the RM can be expressed as 
%
\begin{eqnarray}
\delta f_{tx}^{\rm RM}(x) & \approx &
	-\frac{D_{x}^{\rm L}}{1 + \kappa_{\alpha}^{-} \xi}
	\left[
	\frac{1}{8} \left(6 + i 7 \xi \alpha_{\rm R} \right) 
	e^{-2 d_{\rm m} / \xi} e^{\kappa_{\alpha}^{+} \left(x-d_{\rm f} \right) } 
	-
	i \frac{1}{2 \xi \alpha_{\rm R}} \left( 1- i \xi \alpha_{\rm R} \right) e^{-\kappa_{\alpha}^{-} \left(x-d_{\rm f} \right) } 
	\right] \nonumber \\
	&\times&
	e^{-d_{\rm f}/\xi} e^{2\left(x-L_{\rm m} \right)/\xi} 
	\left(
	\frac{\Delta_{\rm R}}{\hbar \omega}
	\right)^{2}
	\frac{\Delta_{\rm L}^{*}}{\hbar \omega}
\label{dfx}, 
\end{eqnarray}
and 
\begin{eqnarray}
\delta f_{tz}^{\rm RM}(x) & \approx &
	i\frac{D_{x}^{\rm L}}{1 + \kappa_{\alpha}^{-} \xi}
	\left[
	\frac{1}{4} \left(1 + i \right) 
	e^{-2 d_{\rm m} / \xi} e^{\kappa_{\alpha}^{+} \left(x-d_{\rm f} \right) } 
	+
	i \frac{1}{2 \xi \alpha_{\rm R}} \left( 1- i 2 \right) e^{-\kappa_{\alpha}^{-} \left(x-d_{\rm f} \right) } 
	\right] \nonumber \\
	&\times&
	e^{-d_{\rm f}/\xi} e^{2\left(x-L_{\rm m} \right)/\xi} 
	\left(
	\frac{\Delta_{\rm R}}{\hbar \omega}
	\right)^{2}
	\frac{\Delta_{\rm L}^{*}}{\hbar \omega}
\label{dfz}.
\end{eqnarray}
Within the present approximation, $\delta f_s(x)$ and $\delta f_{ty}(x)$ vanish.
These results indicate that the SHJC is solely carried by STCs.
In the next subsection, we calculate the SHJC using Eqs.~(\ref{dfx}) and (\ref{dfz}).

\subsection{Formulation of the second harmonic Josephson current}
In this subsection, we derive the SHJC using Eqs.~(\ref{f-apx}), (\ref{dfx}), and (\ref{dfz}). 
The Josephson current in the quasiclassical theory near the superconducting transition temperature is expressed as 
\begin{eqnarray}
j_{\rm T}(x) &=& i\frac{e \pi D N_{\rm F} k_{\rm B} T}{2}
	{\rm tr} 
	\left[
	{\hat F}(x) \tilde \partial _x {\hat F}^{\dagger}(x) 
	-
	{\hat F}^{\dagger}(x) \tilde \partial _x {\hat F}(x)
	\right] \nonumber \\
	&\approx& 
	i\frac{e \pi D N_{\rm F} k_{\rm B} T}{2}
	{\rm tr} 
	\left[
	\left(
	{\hat f}(x) +\delta {\hat f}(x)
	\right)
	\tilde \partial _x 
	\left(
	{\hat f}^{\dagger}(x) + \delta {\hat f}^{\dagger}(x) 
	\right)
	-
	\left(
	{\hat f}^{\dagger}(x) + \delta {\hat f}^{\dagger}(x) 
	\right)
	\tilde \partial _x 
	\left(
	{\hat f}(x) +\delta {\hat f}(x)
	\right)
	\right] \nonumber \\
	&\approx& j_{\rm Q}(x) + \delta j_{\rm Q}(x)
\label{ix}, 
\end{eqnarray}
where $j_{\rm Q}(x)$ is the FHJC given by Eq.~(\ref{jf}) and $\delta j_{\rm Q}(x)$ is the SHJC. 
$\delta j_{\rm Q}(x)$ by considering $\delta f_{s(ty)}(x)=0$ is expressed as 
\begin{eqnarray}
\delta j_{\rm Q}(x) &=& \delta j_{1}(x) + \delta j_{2}(x), \\
\delta j_{1}(x) &=& 
	2 e \pi D N_{\rm F} k_{\rm B} T
	{\rm Im}
	\left[
	f_{tx}(x) \partial _x \delta f_{tx}^{\dagger}(x)
	+
	f_{tz}(x) \partial _x \delta f_{tz}^{\dagger}(x)
	+
	\delta f_{tx}(x) \partial _x f_{tx}^{\dagger}(x)
	+
	\delta f_{tz}(x) \partial _x f_{tz}^{\dagger}(x)
	\right]
\label{djq1}, \nonumber \\
\end{eqnarray}
and 
\begin{eqnarray}
\delta j_{2}(x) &=&
	-8 \alpha_{\rm R} e \pi D N_{\rm F} k_{\rm B} T
	{\rm Im}
	\left[
	f_{tx}(x) \delta f_{tz}^{\dagger}(x) - f_{tz}(x) \delta f_{tx}^{\dagger}(x)
	\right]
\label{dja2}, 
\end{eqnarray}
Substituting $x=L_{\rm m}$ into Eqs.~(\ref{fx-rm}), (\ref{fz-rm}), (\ref{dfx}), and (\ref{dfz}) 
and inserting the results into Eqs.~(\ref{djq1}) and (\ref{dja2}), the SHJC as a function of $d_{\rm f}$ and $d_{\rm m}$ is represented by 
\begin{eqnarray}
\delta j_{1}(L_{\rm m}) &\approx&
	\frac{e \pi D N_{\rm F} k_{\rm B} T }{2 \xi} \left(D_{x}^{\rm L} \right)^{2}
	\left(
	\frac{\Delta}{\hbar \omega}
	\right)^{4}
	\delta A_{\omega}(L_{\rm})
	e^{-2 L_{\rm m}/\xi} \sin \left(2\theta \right) \nonumber \\
	&+&
	\frac{e \pi D N_{\rm F} k_{\rm B} T }{2 \xi} \left(D_{x}^{\rm L} \right)^{2}
	\left(
	\frac{\Delta}{\hbar \omega}
	\right)^{4}
	\delta B_{\omega}(L_{\rm m})
	e^{-2 L_{\rm m}/\xi} \cos \left(2\theta \right)
\label{djq2}, \\
	\delta A_{\omega}(L_{\rm}) &=& 
	\frac{5}{2} \sin \left(4 \alpha_{\rm R} d_{\rm m} \right)
	+
	\frac{1}{\xi \alpha_{\rm R}}
	\left[
	\cos \left(4 \alpha_{\rm R} d_{\rm m} \right) 
	+3\sin\left(4 \alpha_{\rm R} d_{\rm m} \right)
	\right], \nonumber \\
	\delta B_{\omega}(L_{\rm m}) &=&
	\frac{5}{2} \sin \left(4 \alpha_{\rm R} d_{\rm m} \right)
	+
	\frac{1}{\xi \alpha_{\rm R}}
	\left[
	\sin \left(4 \alpha_{\rm R} d_{\rm m} \right) 
	-3\cos\left(4 \alpha_{\rm R} d_{\rm m} \right)
	\right], \nonumber	
\end{eqnarray}
where $\delta j_{2}(L_{\rm m}) = 0$ becomes zero in the present calculation. 
From Eq.~(\ref{djq2}), it can be clearly seen that a $\varphi_{0}$ phase shift can emerge 
due to an additional cosine term in the second harmonic Josephson current, even in the absence of an external magnetic field.
Consequently, owing to the $\varphi_{0}$ phase shift in both the FHJC and the SHJC, the JDE can occur without requiring the external magnetic field.
%
\begin{figure}[!t]
\begin{center}
\includegraphics[width=12cm]{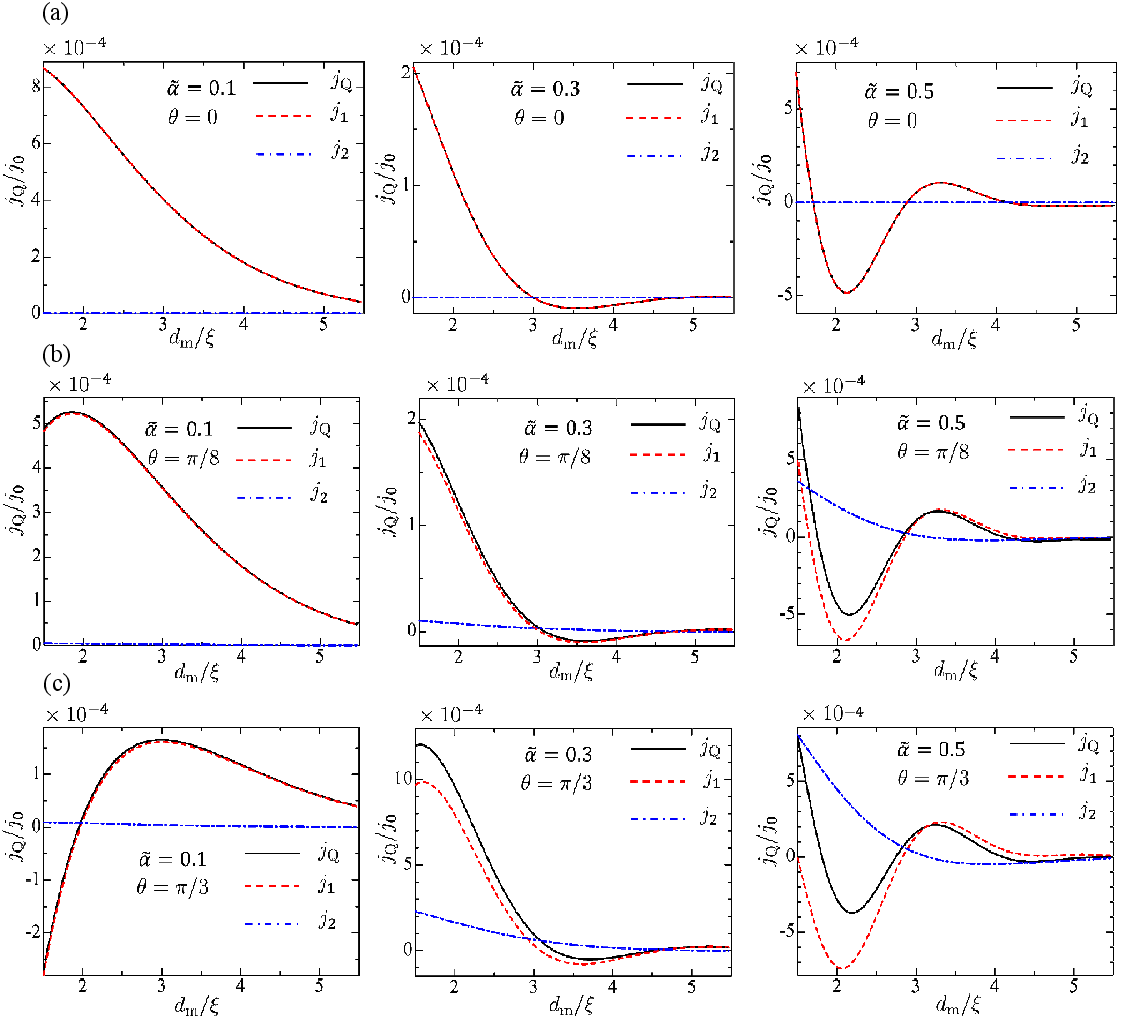}
\caption{ (Color online) 
The first harmonic Josephson current as a function of the thickness of the RM ($d_{\rm m}$) for $\tilde{\alpha} = \xi \alpha_{\rm R} =$ 0.1, 0.3, and 0.5.
In Figs.~(a)–(c), we used $\theta$ = 0, $\pi/8$, and $\pi/3$, respectively. 
Here, $j_{\rm Q}(L_{\rm m}) = j_{1}(L_{\rm m}) + j_{2}(L_{\rm m})$, $\xi_{(\rm C)}=\sqrt{\hbar D/2 \pi k_{\rm B}T_{(\rm C)}}$, and $j_{0}=e N D \Delta_{0}/\pi k_{\rm B} T_{\rm C} \xi_{\rm C}$. 
$\xi_{\rm C}$ denotes $\xi$ evaluated at $T = T_{\rm C}$.
}
\label{jq-d}
\end{center}
\end{figure}
%
\begin{figure}[t]
\begin{center}
\includegraphics[width=6cm]{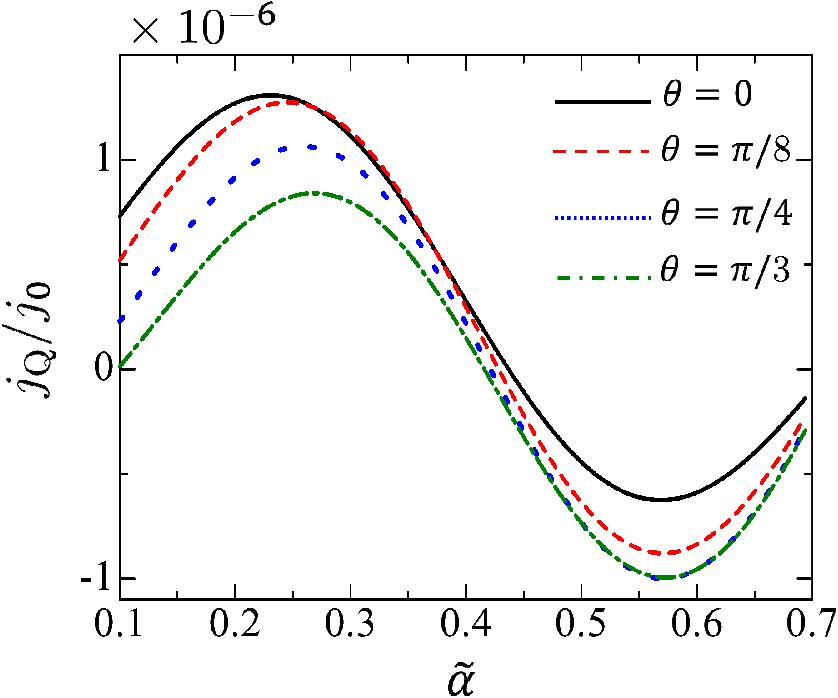}
\caption{ (Color online) 
The first harmonic Josephson current as a function of the RSOI $\tilde{\alpha} = \xi \alpha_{\rm R} $ for $\theta$ = 0, $\pi/8$, $\pi/4$, and $\pi/3$.
Here, $j_{\rm Q}(L_{\rm m}) = j_{1}(L_{\rm m}) + j_{2}(L_{\rm m})$, $\xi_{(\rm C)}=\sqrt{\hbar D/2 \pi k_{\rm B}T_{(\rm C)}}$, and $j_{0}=e N D \Delta_{0}/\pi k_{\rm B} T_{\rm C} \xi_{\rm C}$. 
$\xi_{\rm C}$ denotes $\xi$ evaluated at $T = T_{\rm C}$.
}
\label{jq-a}
\end{center}
\end{figure}
\section{Numerical results}
In this section, we show numerical results of the FHJC and the SHJC using Eqs.~(\ref{jq}), (\ref{ja}), and (\ref{djq2}). 
In numerical calculations, the temperature dependence of the superconducting gap $\Delta$ is assumed to be $\Delta=1.74 \Delta_{0} \left( 1 - T/T_{\rm C} \right)^{1/2}$, 
where $\Delta_{0}$ and $T_{\rm C}$ are the superconducting gap at zero temperature and the superconducting transition temperature, respectively\cite{tinkham-book}. 
In all numerical calculations, the normalized temperature is set to $T/T_{\rm C} = 0.9$; 
the polar angle of the magnetization in the ${\rm F}_{\rm L(R)}$ are fixed at $\theta_{\rm L(R)} = \pi/4$; 
and the thicknesses are set to $d_{\rm L(R)}/\xi_{\rm f} = 0.4$ and $d_{\rm f}/\xi = 2$. 
Here, $\xi_{\rm f} = \sqrt{\hbar D/h}$ is the characteristic penetration length of Cooper pairs in the ferromagnetic region.
In all figures, the variable $L_{\rm m}$ for the Josephson current is omitted. 
Based on the above, qualitative behavior of the FHJC and the SHJC is provided below. 

\subsection{First harmonic Josephson current}
Figure 2 shows the numerical results of the FHJC, $j_{\rm Q}(L_{\rm m})$, 
as a function of the thickness of the RM, $d_{\rm m}$, obtained using Eqs.~(\ref{jq}) and (\ref{ja}). 
It is clearly observed that $j_{\rm Q}(L_{\rm m})$ exhibits damped oscillatory behavior as a function of $d_{\rm m}$ 
with increasing $\alpha_{\rm R}$ as shown in Figs.~\ref{jq-d} (a) -- (c). 
Due to its damped oscillatory behavior, $j_{\rm Q}(L_{\rm m})$ changes sign, i.e., 
it is possible to reverse the direction of the Josephson current as $d_{\rm m}$ varies. 
The oscillation period of $j_{\rm Q}(L_{\rm m})$ decreases with increasing $\alpha_{\rm R}$. 
Moreover, $j_{2}(L_{\rm m})$, which is nonzero for finite $\alpha_{\rm R}$, increases with $\theta$.
Therefore, $j_{2}(L_{\rm m})$ provides a significant contribution to the Josephson current. 

\begin{figure}[!t]
\begin{center}
\vspace{30 mm} 
\includegraphics[width=6cm]{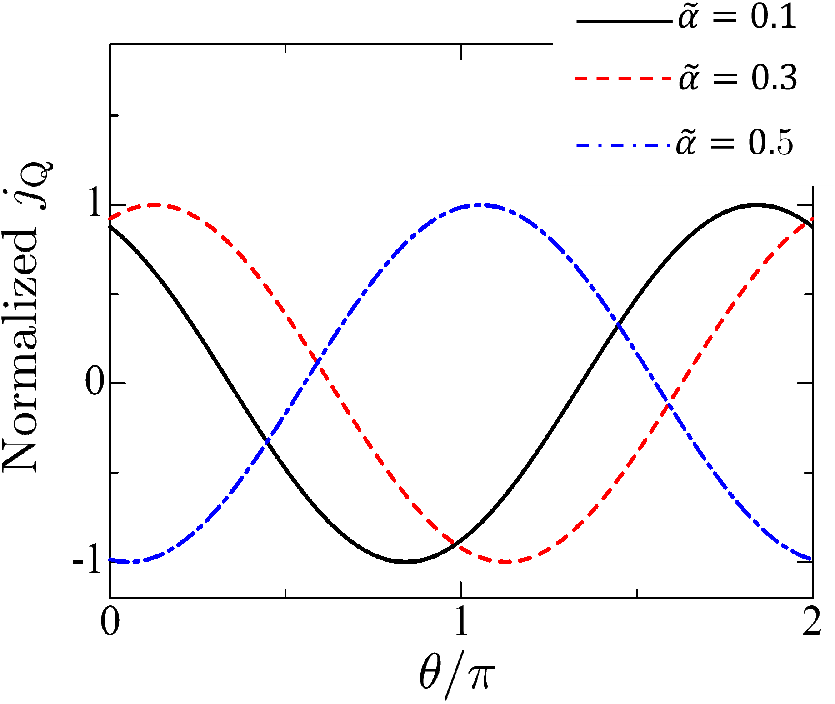}
\caption{ (Color online) 
The current-phase relation in the first harmonic Josephson current for $\tilde{\alpha}$ = 0.1, 0.3, and 0.5. 
Here, $j_{\rm Q}(L_{\rm m}) = j_{1}(L_{\rm m}) + j_{2}(L_{\rm m})$, $\xi_{(\rm C)}=\sqrt{\hbar D/2 \pi k_{\rm B}T_{(\rm C)}}$, and $j_{0}=e N D \Delta_{0}/\pi k_{\rm B} T_{\rm C} \xi_{\rm C}$. 
$\xi_{\rm C}$ denotes $\xi$ evaluated at $T = T_{\rm C}$.
}
\label{jq-q}
\end{center}
\end{figure}
Figure~3 shows the RSOI dependence of the FHJC for $\theta = 0$, $\pi/8$, $\pi/4$, and $\pi/3$.
It is clearly observed that the direction of the FHJC is reversed as $\alpha_{\rm R}$ is varied.
Experimentally, $\alpha_{\rm R}$ can be controlled by tuning the gate voltage, offering a feasible way to switch the direction of the Josephson current.
In addition, the amplitude of the FHJC is found to depend on $\theta$.
$\theta$ can also be tuned experimentally, for example, by applying a DC bias current to the junction.
These observations indicate that both the magnitude and direction of the FHJC can be effectively manipulated via external controls.

Figure~\ref{jq-q} shows the current-phase relation (CPR) in the FHJC for $\tilde{\alpha}=0.1, 0.3$, and 0.5.
In conventional Josephson junctions, the CPR follows a simple sine function with zero initial phase, i.e., $I(\theta) \propto \sin \theta$.
However, in the present Josephson junction studied here, the CPR exhibits a phase-shifted sine function depending on $\tilde \alpha$,
i.e., $I(\theta) \propto \sin (\theta + \varphi_{0})$, where $\varphi_{0}$ is an initial phase independent of the DC bias current.
This phase shift originates from the exchange field and the RSOI, as mentioned in Subsections 2.2 and 2.3.
The phase shift of the Josephson current, as shown in Fig.~\ref{jq-q}, is a key phenomenon for realizing the JDE.

\subsection{Second harmonic Josephson current}
\begin{figure}[t]
\begin{center}
\includegraphics[width=15cm]{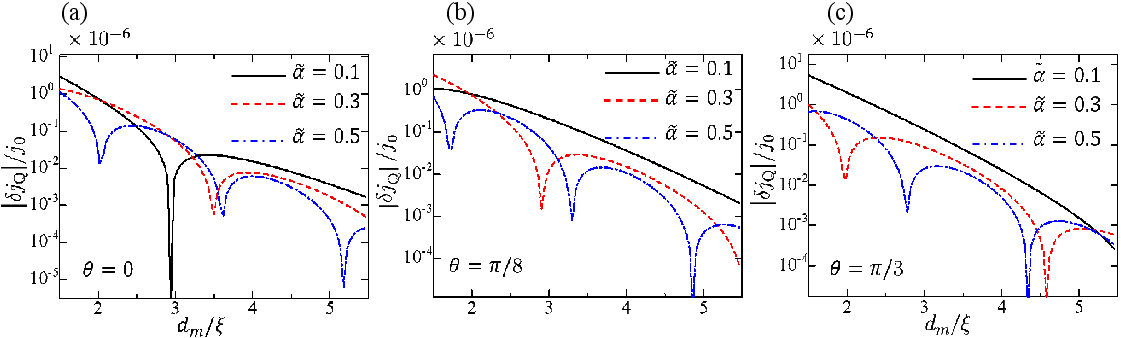}
\caption{ (Color online) 
The second harmonic Josephson current (SHJC) as a function of the thickness of the RM ($d_{\rm m}$) 
for $\tilde{\alpha} = \xi \alpha_{\rm R} =$ 0.1, 0.3, and 0.5.
In Figs.~(a)–(c), we used $\theta$ = 0, $\pi/8$, and $\pi/3$, respectively. 
In this figure, the absolute value of the SHJC is shown. 
Here, $\delta j_{\rm Q}(L_{\rm m}) = \delta j_{1}(L_{\rm m})$, $\xi_{(\rm C)}=\sqrt{\hbar D/2 \pi k_{\rm B}T_{(\rm C)}}$, and $j_{0}=e N D \Delta_{0}/\pi k_{\rm B} T_{\rm C} \xi_{\rm C}$. 
$\xi_{\rm C}$ denotes $\xi$ evaluated at $T = T_{\rm C}$.
}
\label{djq-d}
\end{center}
\end{figure}
Figure~\ref{djq-d} shows the SHJC as a function of $d_{\rm m}$ for $\theta = 0$, $\pi/8$, and $\pi/3$.  
The expression for the SHJC is derived under the condition $\alpha_{\rm R} \neq 0$, as described in Sec.~3.  
It is clearly observed that $\delta j_{\rm Q}(L_{\rm m})$ exhibits damped oscillatory behavior as a function of $d_{\rm m}$, similar to the behavior of the FHJC.  
Notably, $\delta j_{\rm Q}(L_{\rm m})$ also shows a monotonic decrease with increasing $\theta$ as seen in Figs.~\ref{djq-d}(b) and \ref{djq-d}(c).  
From Figs.~\ref{djq-d}(a)--(c), it is evident that the oscillation period of $\delta j_{\rm Q}(L_{\rm m})$ depends on both $\theta$ and $\tilde{\alpha}$.  
Furthermore, the oscillation period becomes shorter as $\tilde{\alpha}$ increases.

\begin{figure}[!t]
\begin{center}
\includegraphics[width=6cm]{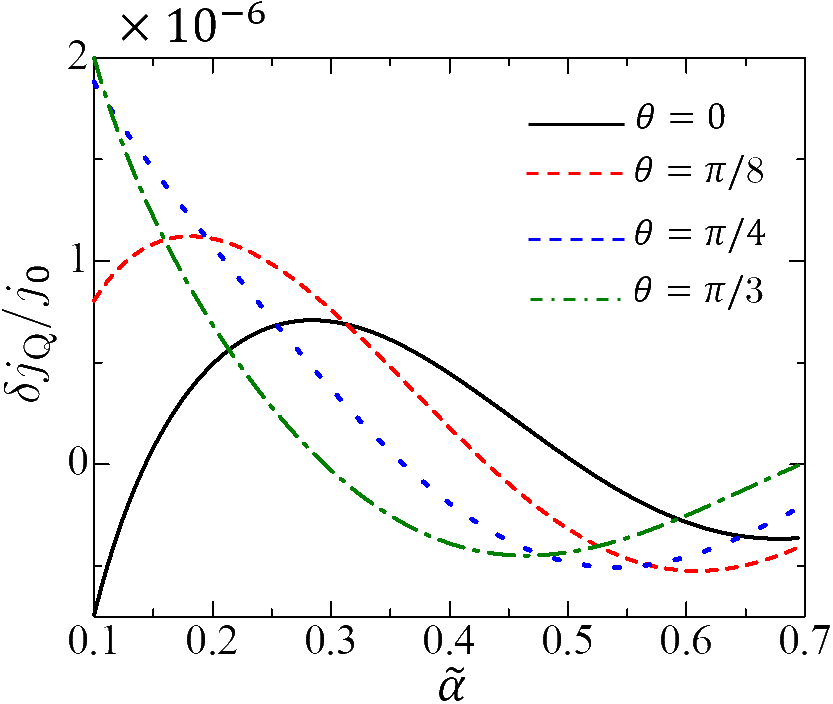}
\caption{ (Color online) 
The Second harmonic Josephson current as a function of the RSOI $\tilde{\alpha} = \xi \alpha_{\rm R} $ for $\theta$ = 0, $\pi/8$, $\pi/4$, and $\pi/3$.
Here, $\delta j_{\rm Q}(L_{\rm m}) = \delta j_{1}(L_{\rm m})$, $\xi_{(\rm C)}=\sqrt{\hbar D/2 \pi k_{\rm B}T_{(\rm C)}}$, 
and $j_{0}=e N D \Delta_{0}/\pi k_{\rm B} T_{\rm C} \xi_{\rm C}$. 
$\xi_{\rm C}$ denotes $\xi$ evaluated at $T = T_{\rm C}$.
}
\label{djq-a}
\end{center}
\end{figure}

Figure~\ref{djq-a} shows the SHJC as a function of $\tilde \alpha$ for $\theta$ = 0, $\pi/8$, $\pi/4$, and $\pi/3$. 
It is possible to change the sign of $\delta j_{\rm Q}(L_{\rm m})$ by tuning $\tilde \alpha$. 
The amplitude of $\delta j_{\rm Q}(L_{\rm m})$ decreases with increasing $\tilde \alpha$. 
Moreover, it is also observed that the the sign of $\delta j_{\rm Q}(L_{\rm m})$ can be controlled by tuning $\theta$. 
Therefore, the direction of $\delta j_{\rm Q}(L_{\rm m})$ can be easily controlled by changing $\theta$ as well as $\tilde \alpha$. 
It should be emphasized that the sign reversal of the SHJC as a function of $\theta$ is a remarkable feature, which is difficult to realize in the FHJC.

%
\begin{figure}[!t]
\begin{center}
\vspace{25 mm} 
\includegraphics[width=6cm]{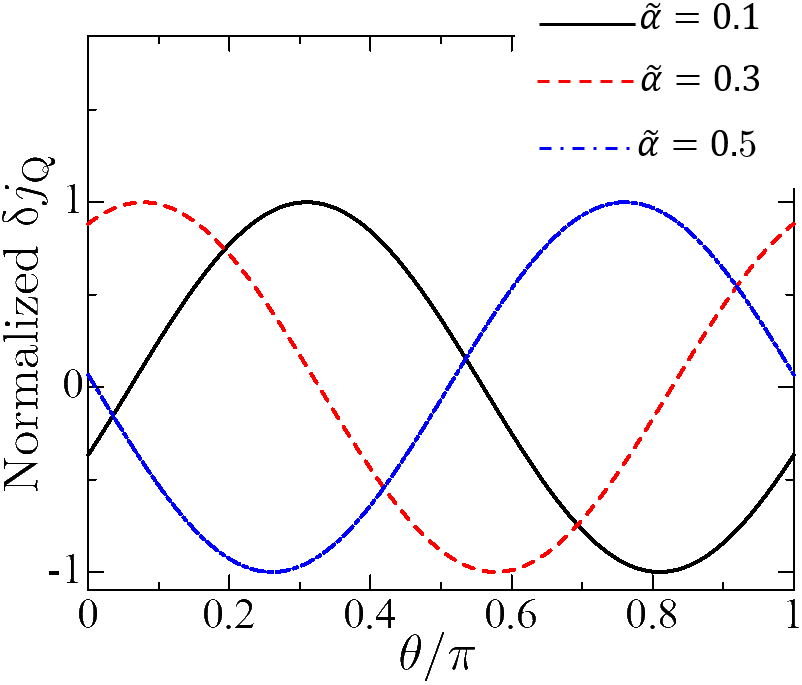}
\caption{ (Color online) 
The current-phase relation in the Second harmonic Josephson current for $\tilde{\alpha} = 0.1, 0.3,$ and $0.5 $. 
Here, $\delta j_{\rm Q}(L_{\rm m}) = \delta j_{1}(L_{\rm m})$.
}
\label{djq-q}
\end{center}
\end{figure}
Figure~\ref{djq-q} shows the CPR of the SHJC for $\tilde{\alpha} = 0.1$, $0.3$, and $0.5$.
In conventional Josephson junctions, the CPR of the SHJC is expressed as $I_{2}(\theta) \propto \sin (2\theta)$.
However, as shown in Fig.~\ref{djq-q}, it is clearly observed that the CPR of the SHJC exhibits a phase shift,
which can be expressed as $I_{2}(\theta) \propto \sin (2\theta + \varphi_{0}')$.
Moreover, it is found that the additional phase $\varphi_{0}'$, which is independent of the DC bias current, depends on $\tilde{\alpha}$.
Similar to $\varphi_{0}$ in the FHJC, $\varphi_{0}'$ plays an important role in realizing the JDE.
In the next subsection, we evaluate the efficiency of the JDE in the present Josephson junction.

\subsection{Efficiency of Josephson diode effect}
In this subsection, we numerically show an efficiency of the JDE. 
The expression for the Josephson current used to evaluate the JDE is given by 
\begin{eqnarray}
j_{\rm T}(L_{\rm m}) = j_{\rm Q}(L_{\rm m}) + \delta j_{\rm Q}(L_{\rm m}), 
\label{jc}
\end{eqnarray}
where $j_{\rm Q}(L_{\rm m}) = j_{1}(L_{\rm m}) + j_{2}(L_{\rm m})$. 
The terms $j_{1}(L_{\rm m})$ and $j_{2}(L_{\rm m})$ are given by Eq.~(\ref{jq}) and Eq~(\ref{ja}), respectively, 
while $\delta j_{\rm Q}(L_{\rm m})$ is given by Eq.~(\ref{djq2}). 
The efficiency of the JDE is expressed as
\begin{equation}
\eta = \frac{j_{\rm c +} - |j_{\rm c -}|}{j_{\rm c +} + |j_{\rm c -}|}, 
\label{eta}
\end{equation}
where $j_{\rm c +}$ and $j_{\rm c -}$ are the maximum and the minimum values of Eq.~(\ref{jc}), respectively. 
It should be also noticed that $j_{\rm c +}$ and $j_{\rm c -}$ are the maximum and the minimum values in the current-phase relation, respectively. 
\begin{figure}[!t]
\begin{center}
\includegraphics[width=7cm]{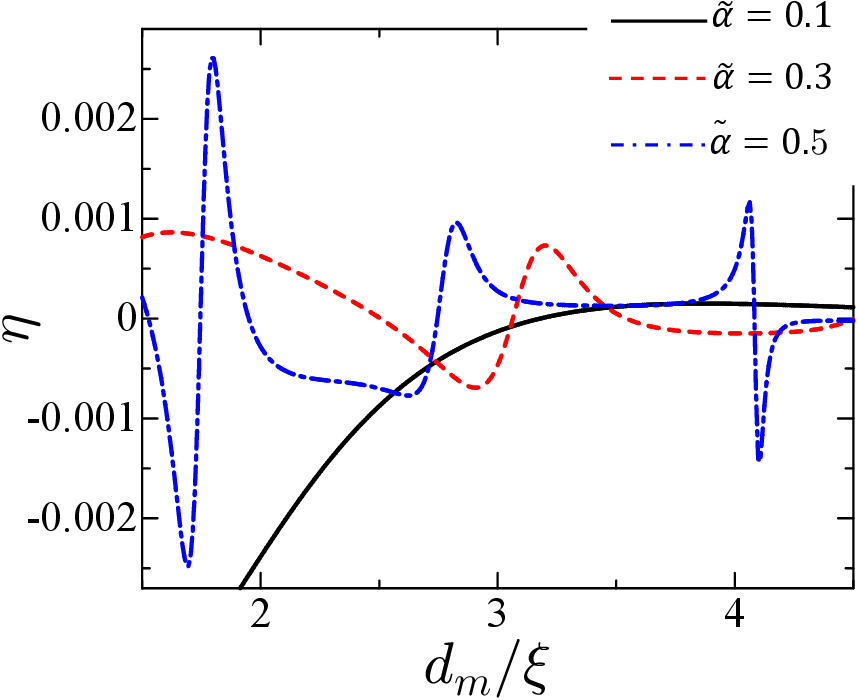}
\caption{ (Color online) 
The efficiency of the Josephson diode effect as a function of $d_{\rm m}$ for $\tilde \alpha$ = 0.1, 0.3, and 0.5. 
Here, $\xi=\sqrt{\hbar D/2 \pi k_{\rm B}T}$. 
}
\label{eta-d}
\end{center}
\end{figure}

Figure~\ref{eta-d} shows the efficiency of the JDE as a function of $d_{\rm m}$ for $\tilde \alpha$=0.1, 0.3, and 0.5. 
In small $\tilde \alpha$, it is found that $\eta$ make smooth changes as a function of $d_{\rm m}$, as shown in the (black) solid line of Fig~\ref{eta-d}. 
With increasing $\tilde \alpha$, it is clearly observed that $\eta$ exhibits a sharp increase at a certain value of $d_{\rm m}$, 
as shown in the (red) dashed and the (blue) dashed-dotted lines of Fig.~\ref{eta-d}. 
The sign of $\eta$ changes from positive to negative and vice versa as $d_{\rm m}$ varies. 
It is also found that the magnitude of $\eta$ decreases with increasing $d_{\rm m}$. 
Therefore, within the present approximation, it is considered that a smaller $d_{\rm m}$ is required to obtain a large $\eta$.

%
\begin{figure}[!h]
\begin{center}
\vspace{20 mm} 
\includegraphics[width=7cm]{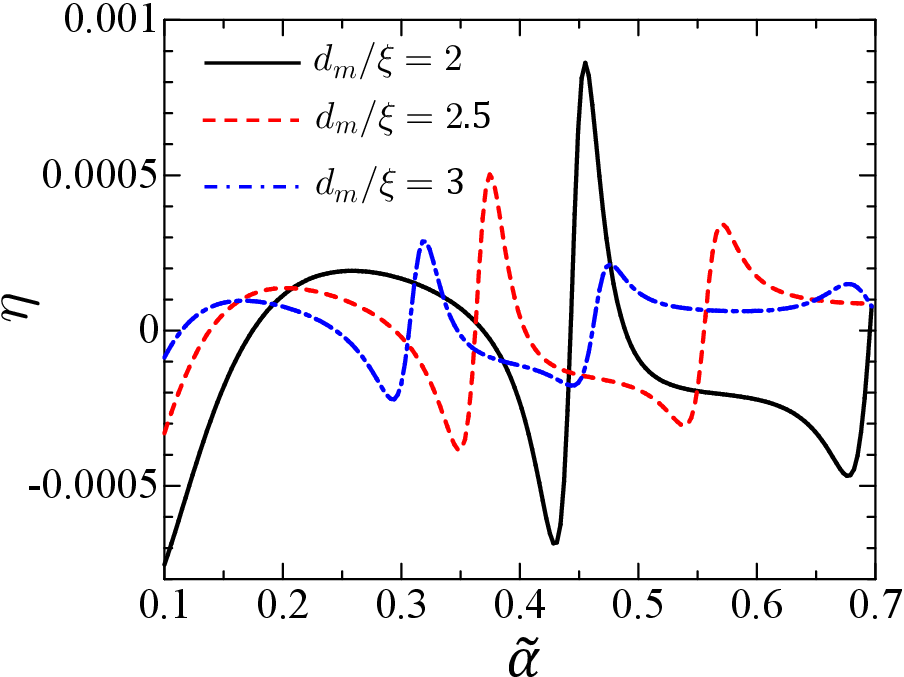}
\caption{ (Color online) 
The efficiency of the Josephson diode effect as a function of $\tilde \alpha$ for $d_{\rm m}/\xi$ = 2, 2.5, and 3. 
Here, $\xi=\sqrt{\hbar D/2 \pi k_{\rm B}T}$. 
}
\label{eta-a}
\end{center}
\end{figure}
Figure~\ref{eta-a} shows the efficiency of the JDE as a function of $\tilde \alpha$ for $d_{\rm m}/\xi$ = 2, 2.5, and 3. 
It is clearly observed that $\eta$ exhibits peak structures a certain value of $\tilde \alpha$, 
as shown in the (black) solid, the (red) dashed, and the (blue) dashed-dotted lines of Fig~\ref{eta-a}. 
With increasing $d_{\rm m}$, the peak structures gradually shift to smaller $\tilde{\alpha}$ values, accompanied by a decrease in the magnitude of $\eta$.
The sing of $\eta$ changes from positive to negative and vice versa as $\tilde \alpha$ varies. 
From Fig~\ref{eta-a}, the large value of $\eta$ can be also obtained by tuning $\tilde \alpha$ as well as $d_{\rm m}$. 

\section{Discussion}

In this section, we discuss possible strategies to enhance the efficiency of the JDE, characterized by the parameter $\eta$, and highlight several future challenges.

The Josephson current is formulated under the assumption that the temperature is close to the superconducting transition temperature. 
In general, the Josephson current becomes significantly larger at temperatures well below the critical temperature. 
Therefore, although we work within this high-temperature approximation, 
it is expected that the efficiency $\eta$ of the JDE could be further enhanced at lower temperatures due to the increased magnitude of the Josephson current.

Another approximation employed in this study is $d_{\rm m}/\xi \gg 1$, 
which allows for analytical expressions for the Josephson current. 
Within this regime, the SHJC, which contains the factor $e^{-2d_{\rm m}/\xi}$, decays more rapidly than the FHJC, 
which contains $e^{-d_{\rm m}/\xi}$. 
As a result, the SHJC is more strongly suppressed with increasing $d_{\rm m}$. 
This implies that a high efficiency $\eta$ is likely to be achieved when the RM layer is thin, where the SHJC remains significant.

The SHJC plays a key role in realizing the JDE. 
As discussed above, thin RM layers are advantageous for enhancing the SHJC and, consequently, 
the efficiency $\eta$. 
Moreover, as shown in Figs.~\ref{eta-d} and \ref{eta-a}, $\eta$ exhibits peak structures as a function of both $d_{\rm m}$ and $\alpha_{\rm R}$. 
Therefore, it may be possible to realize a highly efficient Josephson diode by carefully optimizing these parameters.

In addition to these tunable parameters, temperature provides an alternative and experimentally accessible control parameter. 
Our numerical calculations demonstrate that the Josephson currents exhibit sign changes when $d_{\rm m}$ and $\alpha_{\rm R}$ are varied. 
Interestingly, similar sign reversals can also be induced by changing the temperature, 
provided that $d_{\rm m}$ is set near a point where the Josephson currents vanish. 
This is because the oscillation period of the Josephson currents depends on 
$\xi = \sqrt{\hbar D / 2\pi k_{\rm B} T} \propto 1/\sqrt{T}$, which is sensitive to temperature variations. 
Such temperature-induced sign changes may offer a practical route to controlling the direction of the Josephson current.

It should be noted, however, that the present theoretical model assumes ideally transparent interfaces between all layers. 
In practice, achieving high interface transparency in multilayer structures is a significant experimental challenge. 
This issue is particularly critical for higher harmonic components of the Josephson current, such as the SHJC, 
which are sensitive to interface quality. 
Consequently, the experimental observation of the SHJC could be difficult under low-transparency conditions.

Despite these challenges, the present study provides fundamental theoretical insights into how the interplay between spin-orbit coupling and magnetism can generate higher harmonic Josephson currents in the absence of an external magnetic field. 
These findings may serve as valuable design principles for future experimental efforts aimed at realizing nonlinear Josephson effects by optimizing interface transparency and multilayer geometry.

\section{Summary}
We have theoretically formulated the first harmonic Josephson current (FHJC) and the second harmonic Josephson current (SHJC), and evaluated the Josephson diode effect (JDE) in the ${\rm S_{L}}/{\rm F_{L}}/{\rm F}/{\rm RM}/{\rm F_{R}}/{\rm S_{\rm R}}$ junction.
Both the FHJC and SHJC exhibit damped oscillatory behavior as functions of the thickness of the RM layer ($d_{\rm m}$) 
and the Rashba spin-orbit interaction strength ($\alpha_{\rm R}$).
Notably, these oscillations indicate that the current direction can be reversed by tuning $d_{\rm m}$ or $\alpha_{\rm R}$.

In the present junction setup, the Josephson current is exclusively carried by spin-triplet Cooper pairs (STCs), 
as the spin-singlet components are strongly suppressed by the thick ferromagnetic layer.
This clearly demonstrates that STCs can play a dominant role in mediating nonreciprocal superconducting transport.

Furthermore, the current-phase relations of both FHJC and SHJC exhibit an additional phase shift, even in the absence of an external magnetic field, revealing the characteristics of the $\varphi_{0}$-junction.
Unlike the conventional superconducting phase difference, this additional phase shift is independent of the DC bias current.
Our analysis shows that this shift originates from the interplay between the exchange field and Rashba spin-orbit interaction.
As a result, the JDE emerges without requiring an external magnetic field, due to the asymmetry in the current-phase relation.

Moreover, our findings suggest that the efficiency of the JDE can be enhanced by optimizing $d_{\rm m}$ and $\alpha_{\rm R}$.
It is expected that these results provide valuable insights for the design of spin-triplet-based superconducting devices with nonreciprocal transport properties.

\appendix{
\section{Equations for Coefficients Satisfying Boundary Conditions} \label{app:coef}
A system of simultaneous equations for determining the coefficients that satisfy the boundary conditions 
is given by
\begin{eqnarray}
C e^{\kappa_{+}d_{\rm f}} + D e^{-\kappa_{+} d_{\rm f}} + E e^{\kappa_{-} d_{\rm f}} + F e^{-\kappa_{-} d_{\rm f}} &=&
M e^{ d_{\rm f}/\xi} + N e^{-d_{\rm f}/\xi}, \nonumber \\
C \xi \kappa_{+} e^{\kappa_{+} d_{\rm f}} - D \xi \kappa_{+} e^{-\kappa_{+} d_{\rm f}} + E \xi \kappa_{-} e^{\kappa_{-} d_{\rm f}} - F \xi \kappa_{-} e^{-\kappa_{-} d_{\rm f}} &=&
M e^{ d_{\rm f}/\xi} - N e^{-d_{\rm f}/\xi}, \nonumber \\
-C e^{\kappa_{+} d_{\rm f}} - D e^{-\kappa_{+} d_{\rm f}} + E e^{\kappa_{-} d_{\rm f}} + F e^{-\kappa_{-} d_{\rm f}} &=&
i G e^{\kappa_{\alpha}^{+} d_{\rm f}} + i H e^{-\kappa_{\alpha}^{-} d_{\rm f}}
+ i I e^{\kappa_{\alpha}^{-} d_{\rm f}} + i J e^{-\kappa_{\alpha}^{+} d_{\rm f}}, \nonumber \\
-C \kappa_{+} e^{\kappa_{+} d_{\rm f}} + D \kappa_{+} e^{-\kappa_{+} d_{\rm f}} + E \kappa_{-} e^{\kappa_{-} d_{\rm f}} - F \kappa_{-} e^{-\kappa_{-} d_{\rm f}} &=&
i G \kappa_{\alpha}^{+} e^{\kappa_{\alpha}^{+} d_{\rm f}} - i H \kappa_{\alpha}^{-} e^{-\kappa_{\alpha}^{-} d_{\rm f}} \nonumber \\
&&+ i I \kappa_{\alpha}^{-} e^{\kappa_{\alpha}^{-} d_{\rm f}} - i J \kappa_{\alpha}^{+} e^{-\kappa_{\alpha}^{+} d_{\rm f}}, \nonumber \\
- i G e^{\kappa_{\alpha}^{+} d_{\rm f}} - i H e^{-\kappa_{\alpha}^{-} d_{\rm f}} 
+ i I e^{\kappa_{\alpha}^{-} d_{\rm f}} + i J e^{-\kappa_{\alpha}^{+} d_{\rm f}}
&=& A e^{ d_{\rm f}/\xi} + B e^{-d_{\rm f}/\xi}, \nonumber \\
- i G \kappa_{\alpha}^{+} e^{\kappa_{\alpha}^{+} d_{\rm f}} + i H \kappa_{\alpha}^{-} e^{-\kappa_{\alpha}^{-} d_{\rm f}} 
+ i I \kappa_{\alpha}^{-} e^{\kappa_{\alpha}^{-} d_{\rm f}} - i J \kappa_{\alpha}^{+} e^{-\kappa_{\alpha}^{+} d_{\rm f}}
&=& A \frac{1}{\xi} e^{ d_{\rm f}/\xi} - B \frac{1}{\xi} e^{-d_{\rm f}/\xi} , \nonumber \\
A e^{d_{\rm L}/\xi} + B e^{-d_{\rm L}/\xi} &=& -\frac{d_{\rm L}^{2} h_{x}^{\rm L}}{\hbar D} \frac{\Delta_{\rm L}}{\hbar \omega}, \nonumber \\
-C\left( 1- \kappa_{+}d_{\rm L} \right) e^{\kappa_{+}d_{\rm L}} 
- D \left( 1+ \kappa_{+}d_{\rm L} \right) e^{-\kappa_{+}d_{\rm L}} \nonumber \\
+E\left( 1- \kappa_{-}d_{\rm L} \right) e^{\kappa_{-}d_{\rm L}} 
+ F\left( 1+ \kappa_{-}d_{\rm L} \right) e^{-\kappa_{-}d_{\rm L}} 
&=&
\frac{d_{\rm L}^{2}h_{z}^{\rm L}}{\hbar D} \frac{\Delta_{\rm L}}{\hbar \omega}, \nonumber \\
C\left( 1- \kappa_{+}d_{\rm L} \right) e^{\kappa_{+}d_{\rm L}} 
+ D \left( 1+ \kappa_{+}d_{\rm L} \right) e^{-\kappa_{+}d_{\rm L}} \nonumber \\
+E\left( 1- \kappa_{-}d_{\rm L} \right) e^{\kappa_{-}d_{\rm L}} 
+ F\left( 1+ \kappa_{-}d_{\rm L} \right) e^{-\kappa_{-}d_{\rm L}} 
&=&
i \frac{\Delta_{\rm L}}{\hbar \omega}, \nonumber \\
-iG \left( 1 + i 2 \alpha_{\rm R} d_{\rm R} \right) e^{\kappa_{\alpha}^{+} L_{\rm m}} 
-iH \left( 1 + i 2 \alpha_{\rm R} d_{\rm R} \right) e^{-\kappa_{\alpha}^{-} L_{\rm m}} \nonumber \\
+ i I  \left( 1 - i 2 \alpha_{\rm R} d_{\rm R} \right) e^{\kappa_{\alpha}^{-} L_{\rm m}} 
+iJ \left( 1 - i 2 \alpha_{\rm R} d_{\rm R} \right) e^{-\kappa_{\alpha}^{+} L_{\rm m}} 
&=&
- \frac{d_{\rm R}^{2}h_{x}^{\rm R}}{\hbar D} \frac{\Delta_{\rm R}}{\hbar \omega}, \nonumber \\
iG \left( 1 + i 2 \alpha_{\rm R} d_{\rm R} \right) e^{\kappa_{\alpha}^{+} L_{\rm m}}  
+iH \left( 1 + i 2 \alpha_{\rm R} d_{\rm R} \right) e^{-\kappa_{\alpha}^{-} L_{\rm m}} \nonumber \\
+I \left( 1 - i 2 \alpha_{\rm R} d_{\rm R} \right) e^{\kappa_{\alpha}^{-} L_{\rm m}} 
+iJ \left( 1 - i 2 \alpha_{\rm R} d_{\rm R} \right) e^{-\kappa_{\alpha}^{+} L_{\rm m}} 
&=&
- \frac{d_{\rm R}^{2}h_{z}^{\rm R}}{\hbar D} \frac{\Delta_{\rm R}}{\hbar \omega}, \nonumber
\end{eqnarray}
and 
\begin{eqnarray}
M e^{L_{\rm m}/\xi} + N e^{-L_{\rm m}/\xi} &=& i \frac{\Delta_{\rm R}}{\hbar \omega}. \nonumber
\end{eqnarray}
}

{}

\end{document}